\documentclass{eas}
\usepackage{graphicx}
\def\vph{\hat{v}_{\phi}}

\begin{document}

\title{On Stellar Dynamo Processes and Differential Rotation} 
\author{Allan Sacha BRUN}\address{JILA, University of Colorado, Boulder, CO 80309-0440, USA.}
\begin{abstract}
Many stars exhibit strong magnetic fields, some of which are thought to 
be of primordial origin and others a sign of magnetic dynamo processes. 
We briefly review the results of observational studies of solar-type stars 
seeking to evaluate the linkage between rotation rate and possible magnetic 
cycles of activity. Clearly turbulent convection and rotation within 
spherical shell geometries provide ingredients essential for dynamo action. 
However, intensive efforts over several decades in solar research have 
demonstrated that it is no easy matter to achieve cyclic magnetic activity 
that is in accord with observations. Helioseismology has revealed that 
an essential element for the global solar dynamo is the presence of a 
tachocline of shear at the base of the solar convection zone, leading 
to the likely operation of an interface dynamo. We review the crucial 
elements for achieving a cyclic magnetic activity. We then discuss some 
of our current 3--D MHD simulations of solar turbulent 
convection in spherical shells that yield differential rotation profiles 
which make good contact with some of the helioseismic findings. We show 
that such turbulent motions can amplify and sustain magnetic field
in the bulk of the convective zone whose strength are sufficient to feed
back both upon the convection and its global circulations.  
\end{abstract}

\runningtitle{A.S. Brun: On Stellar Dynamo Processes and Differential Rotation}
\maketitle
\section{Introduction}
During the last three decades considerable evidence for stellar magnetic activity
has been gathered (Wilson 1978, Noyes et al. 1984, Baliunas et al. 1995). 
These observations based on CaII emissions, revealed that some stars exhibit irregular 
magnetic activity while others have solar-like activity cycles. 
What could be the source of such magnetic fields? Are the different types of activity 
(cyclic vs. irregular) indications of differences in the generation and
amplification of magnetic field? Based on the ohmic diffusion time 
$\tau_{ohm}\sim R^2/\eta$ (where $R$ is the stellar radius and
$\eta$ is the magnetic diffusivity, leading to $\tau_{ohm}\sim 10^{10}$ years for the Sun), 
many of those stars could have retained a fossil magnetic field over their main
sequence lifetime without running a dynamo (Weiss 1994). However, it is 
tempting to search for a link between stellar global properties such 
as mass, rotation or convection and stellar magnetic activity. 
It is indeed important to notice that convection is always present in stars and 
that most of them rotate. Stars less massive than $\sim 1.3$ solar mass $M_{\odot}$ 
display convective envelope whereas the more massive ones possess a convective core. 
The latter appear due to an increasing contribution  to the star's energy generation 
of the CNO nuclear cycles. Further, the range of rotation rates varies significantly 
from very slow rotators ($P > 150$ days) to near breakup velocities. One general trend 
is that older stars rotate slower. Another is that the presence of magnetic fields 
increases the probability for a star to be a slow rotator. More precisely 
Durney \& Latour (1978) have shown that stellar magnetic activity 
for main sequence stars can reasonably been explained by evaluating the  
Rossby number $R_o$. In their model, strong magnetic activity is seen in stars 
for which $R_o << 1$ (i.e. their convective motions strongly feel rotation). 
In reality stars are more complex, and there is evidence that slowly rotating Ap 
stars exhibit stronger magnetic field than fast rotating A stars (Moss 2002).
Here even though A stars have a convective core that is likely to generate a dynamo field, some
quenching mechanism has to be invoked to reproduce the observations--unless the observed field 
in Ap star is of fossil origin and the dynamo generated field screened by 
the radiative envelope. I will here focus my attention on solar type stars 
(see Mestel 2001 for stars of other spectral type). 
In particular, I will examine the dynamical establishment of the solar differential rotation 
and meridional circulations and their nonlinear link with magnetic fields. 
I will present recent results from simulations of solar convection within 
full 3--D spherical shells, discussing the angular velocity $\Omega$ profiles that can be 
achieved in the bulk of such convection zones and the level of dynamo induced magnetism 
that can be sustained there.
 
\section{Our Numerical Approach}

The anelastic spherical harmonics (ASH) code solves the 3--D MHD anelastic equations 
of motion in a rotating spherical shell geometry using a pseudo-spectral/semi-implicit method 
(Miesch et al. 2000, Brun \& Toomre 2002,2003). The model is a simplified description 
of the solar convection zone: solar values are taken for the heat flux, rotation rate, 
mass and radius, and a perfect gas is assumed since the upper boundary of the shell 
lies below the H and He ionization zone. 
The computational domain extends from 0.72 $R_\odot$ to 0.97 $R_\odot$, thereby 
concentrating on the bulk of the unstable zone and here not dealing with penetration 
into the radiative interior. The effects of the steep entropy gradient close to the 
surface has been softened by introducing a subgrid scale (SGS) transport of heat to 
account for the unresolved motions, and enhanced eddy diffusivities are used in 
these large eddy simulations (LES). The typical density difference 
across the shell in radius is about 30.

\section{Solar Convection and Rotation}
\subsection{Convective Patterns}

Figure 1 displays the evolution of enstrophy (vorticity$^2$) in the case called $E$ 
(Brun \& Toomre 2003) over 10 days in time near the top of the domain. 
The vantage point is in the uniformly rotating frame used in our simulations. 
The convection patterns are highly time dependent. Some of the pattern evolution is related to 
the advection by zonal flows associated with the differential rotation driven by the 
convection relative to this frame (i.e fast equator/slow poles). There is an asymmetry 
between the broad upflows showing a weaker vortical content and the narrow fast downflows 
at their periphery. This leads to a downward transport of kinetic energy. 
The strong correlations between warm upward motions and 
cool downward motions are essential in transporting the heat outward. 

\begin{figure}[!ht]
\setlength{\unitlength}{1.0cm}
\begin{picture}(5,4)
\includegraphics{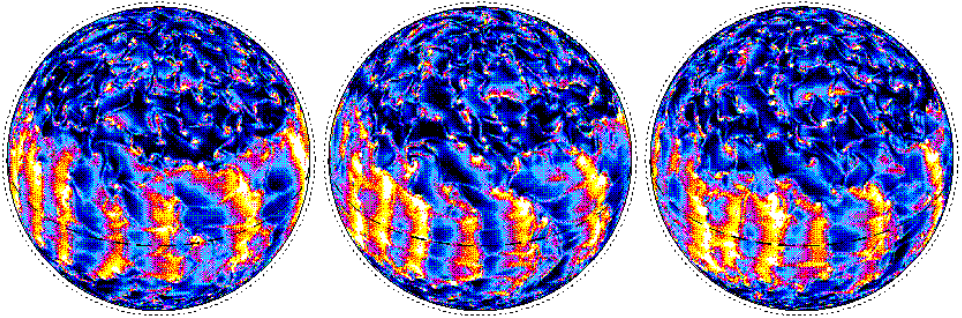}
\end{picture}
\caption[]{\label{fig2} Evolution in the convection over 10 days, showing the 
enstrophy (vorticity$^2$) in case $E$ near the top (0.97 $R_\odot$) of the spherical domain. 
The time interval between each successive image is about 5 days. Strong vortical flows 
appear bright. The dotted circle is located at the solar radius $R_\odot$ and the 
equator is indicated by the dashed curve.}
\end{figure}

Vortical structures are found at the interstices of the downflow network,
although strong fronts are seen in a band near the equator.
The strongest of these vortex tubes or `plumes' extend through the whole domain depth. 
These plumes represent coherent structures that are surrounded by more chaotic flows.
They are counterclockwise in the northern hemisphere and clockwise in the southern one. 
They tend to align with the rotation axis and to be tilted away from the meridional planes,  
leading to Reynolds stresses that are crucial ingredients in redistributing the 
angular momentum within the shell (Brun \& Toomre 2002).

\subsection{Differential Rotation} 

\begin{figure}[!ht]
\setlength{\unitlength}{1.0cm}
\begin{picture}(5,5)
\includegraphics{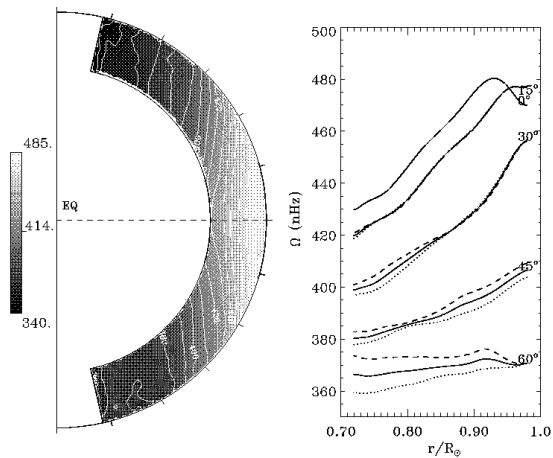}
\end{picture}
\caption[]{\label{fig3} Temporal and longitudinal averages in case $E$ of the angular velocity 
profile formed over an interval of 85 days. This case exhibits a prograde
equatorial rotation and a strong contrast $\Delta \Omega$ from equator to pole, as well as
possessing a high latitudes region of particularly slow rotation. In the right panel a sense
of the asymmetry present in the solution can be assessed in these radial cuts at indicated latitudes.}
\end{figure}

The differential rotation profile in latitude and radius associated with the vigorous 
convection of case $E$ is shown in Figure 2. For simplicity, we have converted the mean 
longitudinal velocity $\vph$ into a sidereal angular velocity $\Omega$, using 
$\Omega_o/2\pi=414$ nHz (or 28 days) as the reference frame rotation rate. 
In the contour plot, the near polar regions have been omitted due to the difficulty 
of forming stable averages there, since the averaging domain is small
but the temporal variations large. Case $E$ exhibits a fast (prograde) equatorial
region and slow (retrograde) high latitude regions. This is due to correlations in the velocity 
components leading to significant Reynolds stresses. These Reynolds stresses are intimately 
linked to the influence of Coriolis forces acting upon the convecting motions and 
to the presence of plumes tilted both away from the local radial direction and out 
of the meridional plane. Such correlations have been identified in local high resolution
Cartesian domains as well (Brummell et al. 1998). These lead to an equatorward transport 
of angular momentum, resulting in the slowing down of the high latitude regions and speeding  
up of the equatorial zone. At low latitudes there is some alignment of $\Omega$ contours 
with the rotation axis. 
At mid latitudes, the angular velocity is nearly constant along radial lines, in good agreement 
with helioseismic deductions (Schou et al 1998). Further, case $E$ exhibits a monotonic decrease 
of $\Omega$ with increasing latitude, a property that has been difficult 
to achieve in 3--D spherical convection calculations. Indeed, most other 
cases have their equator to pole contrast $\Delta\Omega$ confined to mid 
latitudes extending to about $\sim 42^\circ$ (Brun \& Toomre 2002). 
The differential rotation contrast between the equator and 
$60^\circ$ in case $E$ is 110 nHz (or 26\% relative to the frame of reference), 
thus being very close to the 92 nHz (or 22\%) variation observed in the Sun.  
A sense of the asymmetry present in case $E$ can be assessed both in the contour plot 
and in the latitudinal cuts (right panel of Fig. 2), where we have plotted $\Omega$ 
in the north (dotted) and south (dashed) hemispheres, along with their mean. 
The convection itself exhibits some asymmetry between the two hemispheres (cf. Fig. 1), and 
so it is not surprising that the mean flows driven by the convection do the same. These asymmetries
are expected to diminish over longer temporal averages. 
Mean field models of the solar differential rotation (Kichatinov \& R\"udiger 1995) 
have advocated that a thermal wind balance (involving pole to equator temperature
contrasts) could be the cause of the non-cylindrical profile in $\Omega$.
This could come about through the baroclinic nature of the convecting motions yielding some
latitudinal heat fluxes, resulting in the breakdown of the Taylor-Proudman theorem. 
Although it is indeed true that case $E$ exhibits latitudinal 
variation of entropy and temperature fluctuations relative to the mean, these are not
the most dominant players everywhere in the shell. A temperature contrast of few degree K seems
compatible with a $\Delta\Omega/\Omega_o$ of $\sim 30\%$. However, we find that the 
Reynolds stresses are the main agents responsible for the equatorial acceleration 
achieved in our simulations, and thus the solar differential rotation is dynamical in origin. 

\subsection{Meridional Circulation}

The meridional circulation associated with the vigorous convection in case $E$ is 
maintained variously by Coriolis forces acting on the differential rotation, by buoyancy forces, 
by Reynolds stresses and by pressure gradients, and thus can be thought as departures 
from a simple geostrophic balance (Brun \& Toomre 2002). 
The meridional circulation seen in case $E$ exhibits multi-cell structure 
both in latitude and radius, and given the competing processes for its origin, 
it is not straightforward to predict. 
Typical amplitudes for the velocity are of order 25 {\rm m/s}, comparable to local helioseismic 
deductions (Haber et al. 2002). The flow is directed poleward at low latitudes, with return flow 
deeper down. The temporal fluctuations in the meridional circulation are large and thus stable  
time averages are only attained by sampling many rotations. The kinetic energy in the 
differential rotation and in the  convective motions are two orders of magnitude higher 
than that in the meridional circulation (Brun \& Toomre 2002). As a result, small 
fluctuations in the convective motions and differential rotation can lead to major 
variations in the circulation. Some of the helioseismic inferences suggest 
the presence of single cell circulations, which are at odds with our multi-cell patterns. 
However these inferences vary from year to year, and there is recent evidence for double-cell 
structure in the circulations observable in the near-surface shear layer, but only 
in the northern hemisphere as the current solar cycle advances (Haber et al. 2002). 
From a careful analysis of the angular momentum transport in our shell we have deduced 
that the slow pole behavior seen in case $E$ seems to come about from a relatively 
weak meridional circulation at high latitudes. This permits a more efficient extraction 
of angular momentum by the Reynolds stresses from the high latitudes toward the equator 
in yielding the interesting differential rotation profile that is achieved. 

\section{Incorporating some Magnetic Fields}

We now turn to consider the influence of magnetic fields both upon the convection in our 
deep shell and upon the angular velocity profiles that can be maintained.
Early attempts to explain the 22-year solar cycle considered the possibility that the solar 
dynamo operated within the bulk of the convective envelope (Gilman 1983, Glatzmaier 1985), 
but such approaches failed because strong magnetic fields could not be stored efficiently
within the unstable stratification of the convection zone. The dynamo periods were too short 
(of the order of 1 year) and the poloidal fields were found to propagate poleward, at variance with 
observations. More recently, Parker (1993) has proposed an {\it interfacial} dynamo model
seated in the stable tachocline. 
Magnetic field is still generated within the bulk of the convection zone ($\alpha$-effect), 
but is pumped downward into the stable layer via overshooting turbulent plumes, 
to be stretched there into large-scale toroidal structures by the strong shear 
of the tachocline ($\omega$-effect). When the amplification of the toroidal magnetic 
field is great enough, the structures (called magnetic flux tubes in mean field models) 
become magnetically buoyant and rise upward through the convective envelope. 
The strongest of those structures emerge in the photosphere as bipolar magnetic 
arcades, whereas the weaker ones are recycled within the convective zone. 
This leads to the crucial natural cycle of poloidal to toroidal exchange, i.e. 
$B_{pol} \rightarrow B_{tor} \rightarrow B_{pol}$. However, many aspects 
of these essential `dynamo building blocks' remain to be demonstrated through 
nonlinear 3--D calculations. At present it is not 
feasible to simulate self consistently all the processes operating together, and thus 
one needs to concentrate on individual components (magnetic generation, pumping, 
shearing, buoyant rising). One important ingredient in the interfacial dynamo 
scenario is the ability of the convective motions to generate and sustain 
magnetic fields in the bulk of the zone (Cattaneo \& Hughes 2001). 
We thus have evaluated some conditions in 3--D compressible convective shells for which 
such a dynamo threshold can be realized. We further wished to identify the maximum 
nonlinear amplification of the magnetic field that can be sustained by the 
convective motions without destroying the strong angular velocity contrasts previously attained.

\subsection{Determining the Dynamo Threshold in Convective Shells}

We have conducted three MHD simulations (named $M1$, $M2$ and $M3$) started from a
solution not as turbulent as case $E$ but exhibiting a 
similar angular velocity profile $\Omega$. We then introduced a small seed dipolar 
magnetic field and let the simulations proceed. Figure 3 shows the magnetic energy 
evolution for three values of the magnetic diffusivity $\eta$ (i.e. 2, 1.6 \& 1 
$\times 10^{12}$ {\rm cm$^2$/s}). We note that over more than 3000 days 
(corresponding to several ohmic decay times) the two lowest diffusive cases 
$M2$ and $M3$ exhibit a sustained magnetic energy (ME), the levels of which depend on $\eta$. 
The other case $M1$ is clearly decaying, since the rate of generation of magnetic fields  
could not compensate for the rate of destruction by ohmic diffusion. 
The dynamo threshold seems to be around a magnetic Reynolds number 
$R_{\rm em}=v_{\rm rms}D/\eta$ of $\sim 300$ or $\eta\sim 1.6\times 10^{12}$. 
This is about 25\% higher than in a progenitor incompressible simulation (Gilman 1983).

\begin{figure}[!ht]
\setlength{\unitlength}{1.0cm}
\begin{picture}(5,5)
\includegraphics{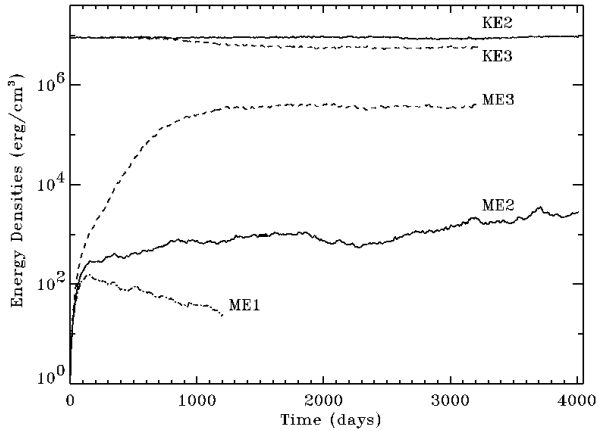}
\end{picture}
\caption[]{\label{fig4} Kinetic energy (KE) and magnetic energy (ME) for cases $M1$, $M2$ and $M3$,
involving in turn a magnetic rms Reynolds number $R_{em}$ of 250 (dashed dot line), 300 
(solid) and 500 (dashed).}
\end{figure}

Further, we find that the kinetic energy (KE) in model $M3$ has been reduced by about 
40\% compared to its initial value. In this case ME has grown to reach a
value of 7\% of KE. The increase in ME started to influence the total amount of 
KE contained in the shell when ME reached roughly 0.5\% of KE in after about 600 days of
evolution. The early exponential growth of ME in case $M3$ corresponds roughly to the 
first 600 days after which the nonlinear feedback of the Lorentz forces on the 
flow begins to be felt. 
For case $M2$, ME is still small enough  (i.e. $\leq 0.1\%$) even after 
4000 days for the convective motions to only be mildly affected by the Lorentz forces. 
This is most clearly seen in comparing the kinetic energy time trace for cases $M2$ and $M3$.
The magnetic energy stored in the toroidal field  within the bulk of the convective zone 
is roughly an order of magnitude greater than that of the poloidal field. 

Figure 4 displays a 3--D rendering of the radial velocity field and of the radial and longitudinal
magnetic fields achieved in our convective shells. The radial component of the magnetic field 
is found to mainly concentrate in the downflow lanes, having been swept away from the center 
of the convective cells by the broad upflows. Both polarities coexist at the downflow network 
interstices. The radial component $B_r$ possesses finer and more intricate structures 
than either $V_r$ or $B_{\phi}$, exhibiting many swirls. 
Substantial magnetic helicity is present, involving complex winding up of the 
toroidal magnetic fields along their length, with both polarities interchanging their positions. 
The toroidal magnetic fields also possess the greater spatial scales, having been stretched 
by the gradients in angular velocity especially in the equatorial region. Some features 
could resemble magnetic flux tubes, although they are short lived and less concentrated.

\begin{figure}[!ht]
\setlength{\unitlength}{1.0cm}
\begin{picture}(10,4.6)
\includegraphics{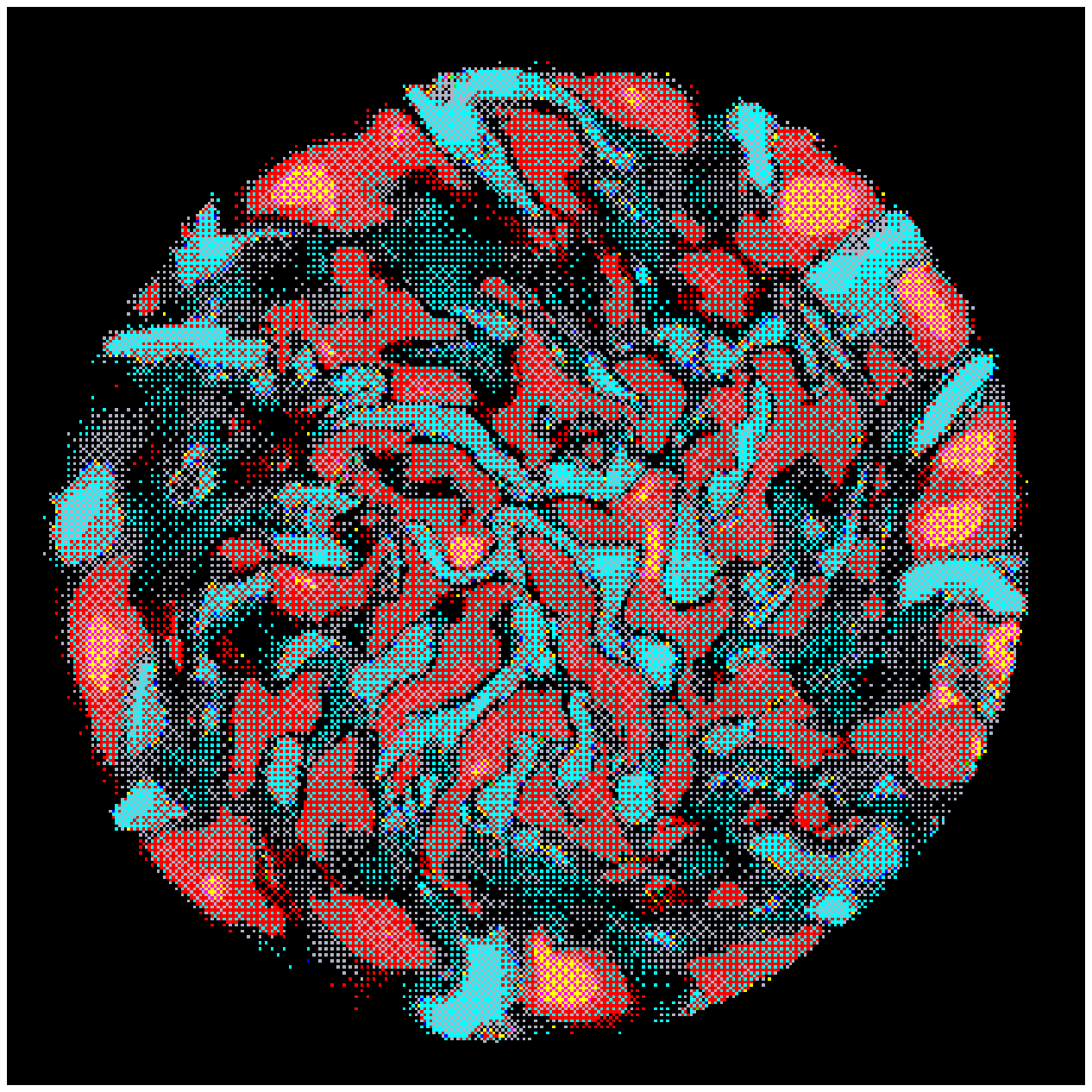}
\includegraphics{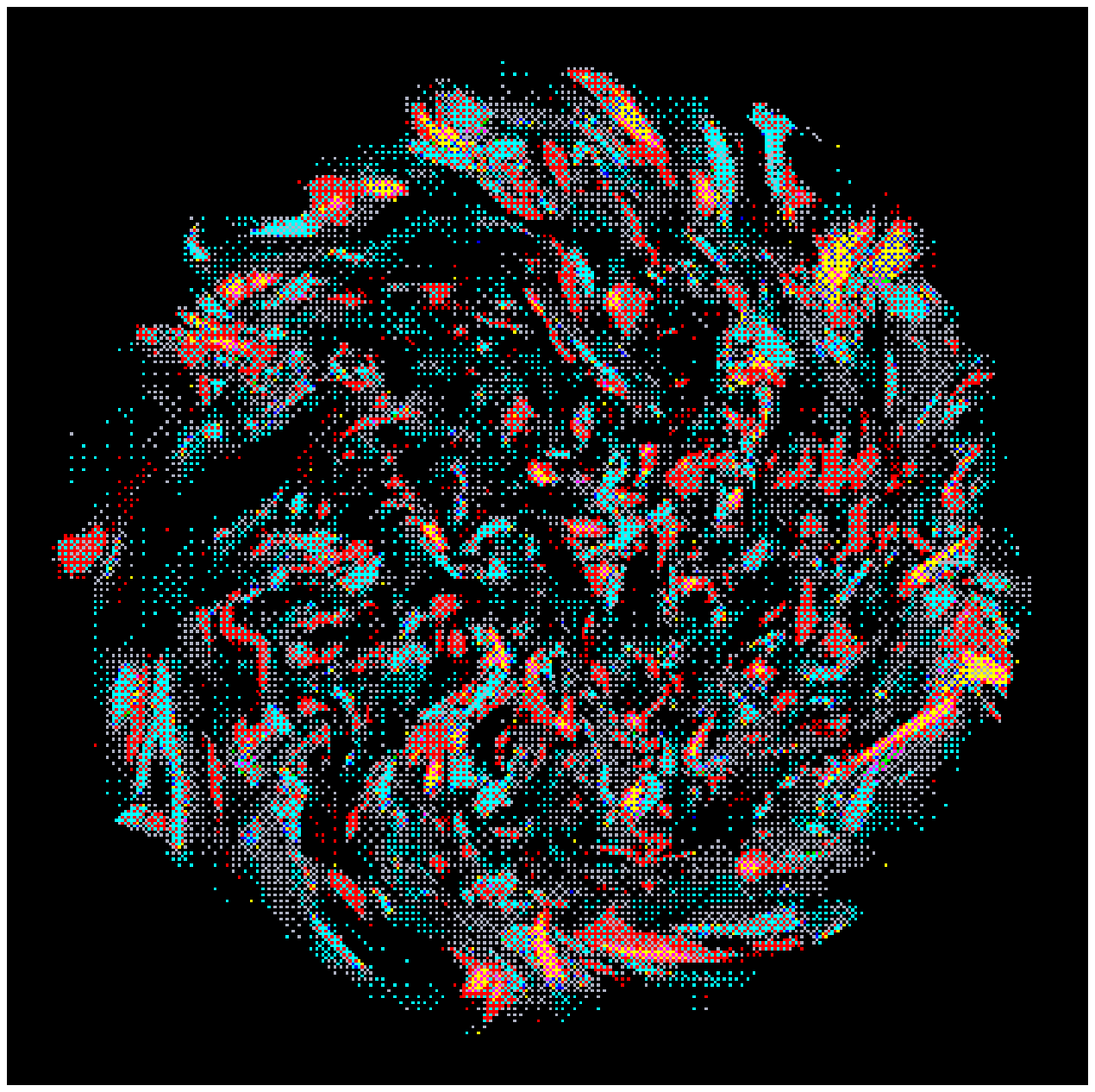}
\includegraphics{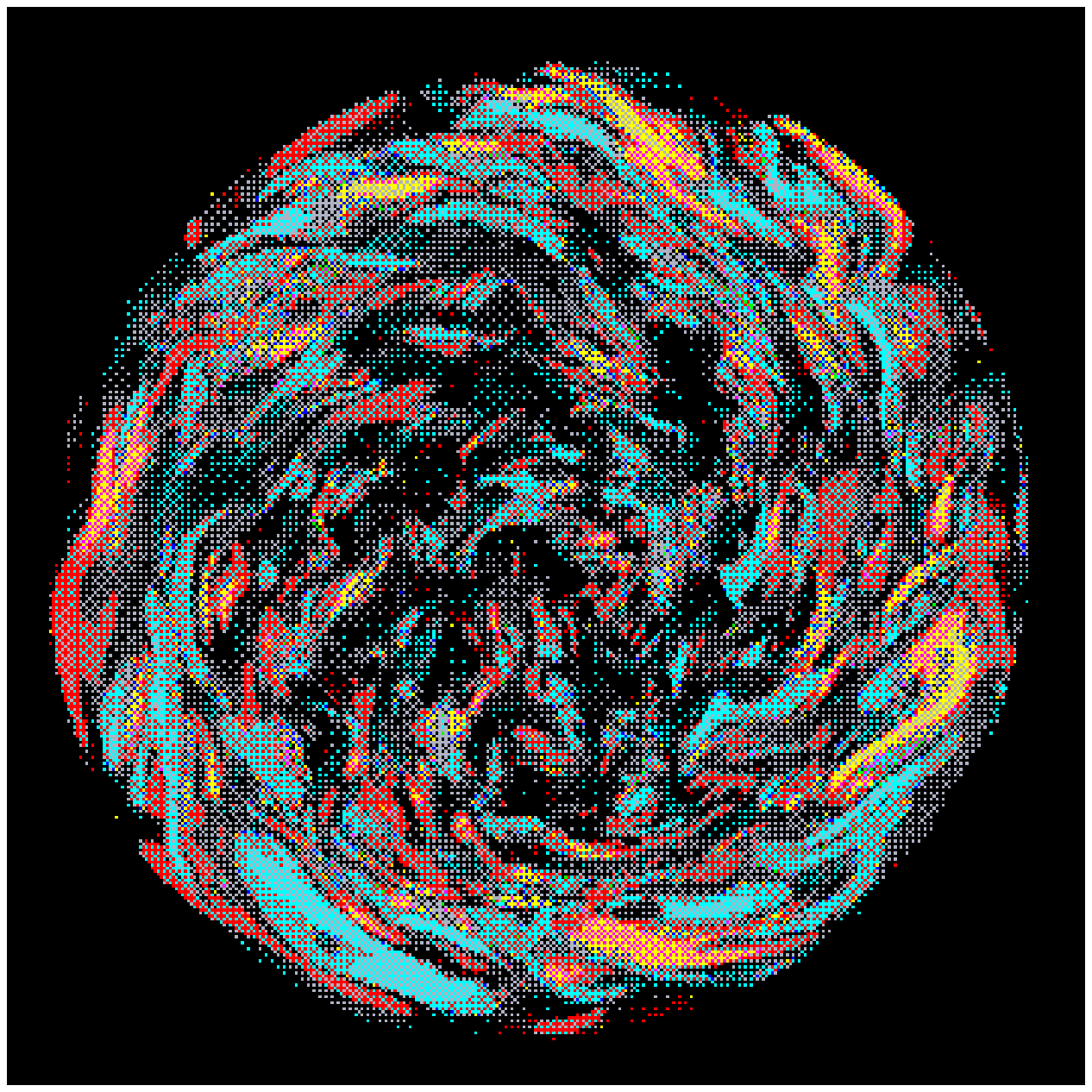}
\end{picture}
\caption[]{\label{fig5} 3-D rendering snapshots of the radial component of the velocity (left) and
of the radial and longitudinal components of the magnetic field (middle and right) in case $M3$. 
Shown is the southern hemisphere seen from within the shell with the equatorial plane facing the reader. 
Brighter tones represent downflow and positive polarity. The velocity and magnetic field 
peak amplitudes are about 100 {\rm m/s} and a few thousand gauss respectively.}
\end{figure}

\subsection{Effect of the Lorentz Force}

With fairly strong magnetic fields sustained within the bulk of the convection 
zone in case $M3$, it is to be expected that the differential rotation $\Omega$ 
will respond to the feedback from the Lorentz forces. Figure 5 shows the time averaged 
angular velocity achieved in case $M3$. The main effect 
of the Lorentz forces is to extract energy from the kinetic energy stored in the 
differential rotation. The reduction of KE contained in the angular velocity is of the
same order as the decrease seen in the total KE, i.e. 40\%. As a consequence,  
the angular velocity contrast $\Delta\Omega$ from 60$^\circ$ to the equator drops 
by $\sim 30\%$ in case $M3$, going from 140 nHz (or 34\% compared to the reference 
frame $\Omega_o$) in the hydrodynamic case to 100 nHz here (or 24\%). Nevertheless,
the angular velocity in case $M3$ remains in reasonable agreement with the solar profile.
The reduction of the latitudinal contrast of $\Omega$ can be attributed 
to the poleward transport of angular momentum by the Maxwell stresses 
(the mean magnetic fields having a negligible contribution). Now the Reynolds stresses 
again need to balance the angular momentum transport by the meridional circulation, 
the viscous diffusion and the Maxwell stresses.
This leads to a less efficient speeding up of the equatorial regions. Since ME is only
7\% of KE in case $M3$, the Maxwell stresses are not yet the main players in
redistributing the angular momentum. We have found that a value of ME above about 20\% of
KE leads to a significant magnetic braking effect on the differential rotation. Had the
simulation been restarted with a stronger initial magnetic field $B_0$, $\Delta\Omega$ could
drop by 90\% in less than a few hundreds days, thus being at variance with helioseismic findings.
Thus the feed back from the Lorentz forces can give an upper limit
on the maximum sustainable magnetic field for a given rotation profile. 
In case $M3$ it seems that the limit is a few thousand gauss but we have to be cautious since this
computation does not allow for expulsion of magnetic helicity or pumping of magnetic field into 
the radiative zone.

\begin{figure}[!ht]
\setlength{\unitlength}{1.0cm}
\begin{picture}(5,5)
\includegraphics{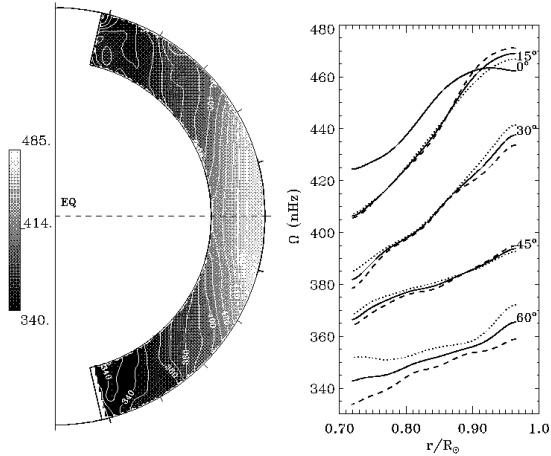}
\end{picture}
\caption[]{\label{fig7} Temporal and longitudinal averages of the angular velocity profiles
achieved in case $M3$ over an interval of 80 days. This case exhibits a prograde
equatorial rotation and a strong contrast $\Delta \Omega$ from equator to pole, as well as
possessing high latitude regions of particularly slow rotation.}
\end{figure}

\subsection{Magnetic Activity and Possible Cycles}

I now wish to discuss possible differences between active stars and
ones with solar-type cycles of magnetic activity. The interfacial dynamo invoked in
this work to explain the solar cycle require a strong rotational shear located in the 
stable tachocline. Without such shear, the star is more likely to undergo local 
magnetic dynamo action directly associated with helical MHD turbulent convective flows 
(i.e the $\alpha$-effect). Yet, any convection zone under the influence 
of rotation is likely to have a more or less pronounced differential rotation 
(cf. Brun, Browning \& Toomre 2003 for core convection and dynamo processes in A-type stars). 
Thus some $\omega$-effect can also be expected within a convection zone, except that the 
strongest magnetic fields present in the unstable convective zone will certainly be smaller 
than the ones amplified in the stable tachocline (again because of buoyancy considerations).
Thus, it is unlikely that purely convective stars such as M dwarfs or pre-main sequence 
stars will exhibit cycles, but instead may display irregular activity (the strength of which will 
likely depends on the rotation rate or Rossby number $R_o$).

\section{Conclusions}

Our simulations of convection in 3-D spherical shells are describing how turbulent
convection under the influence of rotation can achieve a well organized 
differential rotation. Transport of angular momentum by tilted turbulent plumes
generate Reynolds stresses leading to an equatorial acceleration, such as seen in the Sun.
When we introduce a seed magnetic field and let the simulation evolve until equilibration
we find solutions where it is possible to sustain both a magnetic field and a solar-like 
differential rotation. The amount of magnetism that can be sustained in a convective shells
for a given rotation profile gives us some clues on how solar magnetism is working.
Strong energetic event like CME's and the pumping of magnetic energy in the stable radiative
zone certainly plays a significant role in regulating the intricated and coupled interplay of
convection, rotation and magnetism. From a more general stellar point of view, it is clear 
that convection under the influence of rotation and magnetism will lead to significant 
activity but not necessarily to cycles. For the latter, a stable shear region such as the solar 
tachocline seems to be a crucial element in order to close the loop of
magnetic field generation and organization.\\

I thank Mark Miesch, Annick Pouquet, Juri Toomre and Jean-Paul Zahn 
for useful discussions and collaborations and Jean Arnaud for its invitation to the workshop.
This work was partly supported by NASA through SEC Theory Program grant NAG5-8133 and 
by NSF through grant ATM-9731676. The simulations with ASH were carried out with NSF 
PACI support of the San Diego Supercomputer Center (SDSC).  
Much of the analysis of the extensive data sets has been done in the Laboratory for 
Computational Dynamics (LCD) within JILA.


\end{document}